\newif\ifAlpha \Alphafalse

\newif \ifhyperlinks    \hyperlinkstrue

\newif \ifDraft         \Drafttrue
\newif \ifPreprint	\Preprintfalse 
\newif \ifReview	\Reviewtrue
\newif \ifFinal		\Finalfalse

\Draftfalse

\ifFinal
  \Anonfalse
\fi

\documentclass[hyphens,sigconf,10pt]{acmart}
\renewcommand\footnotetextcopyrightpermission[1]{}
\ifAlpha
  \setcitestyle{acmnumeric,super,sort&compress}
  \let \citep = \cite
  \let \citet = \cite
\else
\fi

\usepackage{verbatim}           

\usepackage{xspace}             

\usepackage{balance}            

\usepackage{subfig}
\usepackage{soul}

\setkeys{Gin}{keepaspectratio=true,clip=true,draft=false}
\graphicspath{{./imgs/}}

\ifDraft
  \usepackage{draftcopy}
  \newcommand{\Comment}[1]{\textbf{\textsl{#1}}}
  \newenvironment{LongComment}[1] 
    {\begingroup\par\noindent\slshape \textbf{Begin Comment[#1]}\par}
    {\par\noindent\textbf{End Comment}\endgroup\par}
  \newcommand{\FIXME}[1]{\textbf{\textsl{FIXME: #1}}}
  \newcommand{\TODO}[1]{\textbf{\textsl{TODO: #1}}}
  
\else
  \newcommand{\Comment}[1]{\relax}
  
  \newcommand{\FIXME}[1]{\relax}
  \newcommand{\TODO}[1]{\relax}
  
\fi

\newcommand{\pp}{prime\&\allowbreak probe\xspace}

\newcommand{\calM}{\mathcal{M}}

\newcommand{\stats}[2]{\boldmath$\calM=#2$\,b, $\mathit n=#1$}

\newcommand{\statsmM}[3]{\boldmath$\calM=#2$\,mb, $\calM_0=#3$\,mb, $\mathit n=#1$}

\usepackage{hyperref}
\ifhyperlinks\else   
  \hypersetup{nolinks=true}
\fi

\newcounter{reqCtr}
\newcommand{\reqbox}[2]{\begin{center}
    \refstepcounter{reqCtr}
    \fbox{\parbox{0.9\linewidth}{\textbf{Requirement~\thereqCtr: #1}\\#2}}
  \end{center}}
\newcommand{\reqRef}[1]{\hyperref[#1]{Requirement~\ref*{#1}}}

\newcommand{\code}[1]{\texttt{#1}}
\newcommand{\Obj}[1]{\textsf{#1}\xspace}

\begin{document}
  \sloppy

  \renewcommand{\sectionautorefname}{Section}
  \renewcommand{\subsectionautorefname}{Section}
  \renewcommand{\subsubsectionautorefname}{Section}
  \renewcommand{\appendixautorefname}{Appendix}
  \renewcommand{\Hfootnoteautorefname}{Footnote}
  \newcommand{\Htextbf}[1]{\textbf{\hyperpage{#1}}}

\renewenvironment{abstract}{\section*{Abstract}}{}
  
\title{Time Protection: the Missing OS Abstraction}

\author{Qian Ge}
\affiliation{UNSW Australia and Data61 CSIRO}
\email{qian.ge@data61.csiro.au}

\author{Yuval Yarom}
\affiliation{The University of Adelaide and Data61 CSIRO}
\email{yval@cs.adelaide.edu.au}

\author{Tom Chothia}
\affiliation{University of Birmingham}
\email{T.P.Chothia@cs.bham.ac.uk}

\author{Gernot Heiser}
\affiliation{UNSW Australia and Data61 CSIRO}
\email{gernot@unsw.edu.au}

\settopmatter{printacmref=false}
\settopmatter{printfolios=true}
\maketitle

\begin{abstract}
  Timing channels enable data leakage that threatens
  the security of computer systems,  from cloud platforms to
  smartphones and browsers executing untrusted
  third-party code. Preventing  unauthorised information flow is a core
  duty of the operating system, however, present OSes are unable to
  prevent timing channels. We argue that OSes must provide \emph{time
    protection} in addition to the established memory protection. We
  examine the requirements of time protection, present a design
  and its implementation in the seL4 microkernel, and evaluate its
  efficacy as well as performance overhead on Arm 
  and x86 processors.
\end{abstract}

\section{Introduction}\label{s:intro}

One of the oldest problems in operating systems (OS) research is how
to confine programs so they do not leak information~\citep{Lampson_73}.
To achieve confinement, the operating system needs to control
all of the means of communication that the program can use.
For that purpose, programs are typically grouped into
\emph{security domains}, 
with the operating system exerting its control on cross-domain
communication.

Programs, however, can bypass OS protection
by sending information over channels that are not intended for communication.
Historically, such \emph{covert channels} were explored within
the context of military multi-level-secure systems~\citep{DoD_85:orange}.
Cloud computing, smartphone apps and server-provided JavaScript
executed in browsers mean that we now routinely share
computing platforms with untrusted, potentially malicious, third-party
code.

OSes have traditionally enforced security through
\emph{memory protection}, i.e.\ spatial isolation of security
domains. Recent advances include formal
proof of spatial security enforcement by the seL4 microkernel
\citep{Klein_AEMSKH_14}, including proof of the absence of covert
\emph{storage channels}~\citep{Murray_MBGBSLGK_13}, i.e.\ channels
based on storing information that can be later
loaded~\citep{Schaefer_GLS_77,DoD_85:orange}. Spatial isolation can
thus be considered a solved problem.

The same cannot be said about temporal isolation.
\emph{Timing channels}, and in particular
\emph{microarchitectural channels}~\citep{Ge_YCH_18}, which exploit
timing variations due to shared use of the hardware, remain a
fundamental OS security challenge that has eluded a comprehensive
solution to date. Its importance is highlighted by recent attacks,
including the extraction of encryption keys across
cores~\citep{Irazoqui_ES_15, Liu_YGHL_15} through
\emph{side channels}, i.e.\ without the cooperation of the key owner.

In contrast, covert channels depend on insider help and are
traditionally considered a less significant threat.
However, in the recent Spectre attack~\cite{Kocher_HFGGHHLMPSY_19},
an adversary uses a covert communication channel from speculatively
executed \emph{gadgets} to leak information. This demonstrates that covert
channels pose a real security risk even where no side-channel attack
is known. Furthermore, covert
channel mechanisms bear the risk of being exploitable as a side channel by an
ingenious attacker.

We argue that it is time to take temporal isolation seriously, and
make \emph{time protection} a primary OS duty, just as the established
memory protection.\footnote{Note that we use the term ``OS'' in a
  generic sense, referring to the most privileged software level that
  is responsible for security enforcement. This could refer to a
  hypervisor or the OS of a non-virtualised system.}
Such a design must eliminate, as far as possible, the
sharing of hardware resources that is the underlying cause of timing
channels. The ultimate aim should be to obtain temporal
isolation guarantees comparable to the spatial isolation proofs of
seL4, but for now we focus on a \emph{design} that is suitable for a
verifiable OS kernel, i.e.\ minimal, general and policy-free.

Specifically,  we make the following contributions:
\begin{itemize}
\item We define the requirements for providing
  time protection, enabling confinement in the presence of
  microarchitectural timing channels (\autoref{s:reqts});
\item we introduce a policy-free \emph{kernel clone} operation that
  allows almost perfect partitioning of a system, i.e.\ almost
  completely removing sharing between security domains (\autoref{s:clone});
\item we present an implementation in seL4 (\autoref{s:impl});
\item we show that our
  implementation of time protection is effective, transparently removing timing channels,
  within limitations of present hardware
  (\autoref{s:timing-chnannel-benchmarks});
\item we show that the overhead imposed by these mechanisms is low
  (\autoref{s:system-benchmarks}).
\end{itemize}

\section{Background}\label{s:background}

\subsection{Covert channels and side channels}

A covert channel is an information flow that uses a mechanism not
intended for information transfer \citep{Lampson_73}. Covert channels
therefore may violate the system's security policy, allowing
communication between security domains that should be
isolated. Note that our use of the term (security)
domain is more general than the standard OS term protection
domain (a specific set of access rights). A
security domain consists of one or more protection domains.

There is a traditional distinction between \emph{storage} and
\emph{timing channels}, where exploitation of the latter requires the communicating
domains to have a common notion of time \citep{Wray_91,Schaefer_GLS_77,DoD_85:orange}.
In principle, it is possible to completely eliminate storage channels,
as was done in
the information-flow proof of seL4 \citep{Murray_MBGBSLGK_13}.
\footnote{Specifically, the proof
  shows that no machine state that is touched by the kernel can be
  used as a storage channel, it does not exclude channels through
  state of which the kernel is unaware.} 

Despite recent progress on proving 
upper bounds for the cache side channels of cryptographic 
implementations~\citep{Kopf_MO_12,Doychev_FKMR_13},
proofs of complete elimination of
timing channels in a non-trivial system are beyond current formal
approaches, and measurements are essential for their analysis.

In a narrow sense, a \emph{covert channel} requires collusion between the domains,
one acting as a
\emph{sender} and the other as a \emph{receiver}. Typical cases of
senders are Trojans, i.e.\ trusted code that operates maliciously, or
untrusted code that is being \emph{confined}~\citep{Lampson_73}. 
Due to the collusion, a covert channel represents
a worst case for bandwidth of a channel.

In contrast, a \emph{side channel} has an unwitting sender,
called the \emph{victim}, who, through its normal operation, is
leaking information to an \emph{attacker} acting as the receiver. An
important example is a victim executing in a virtual machine (VM) on
a public cloud, who is being attacked by a malicious co-resident VM~\citep{Yarom_Falkner_14,Inci_GIES_16}.

\subsection{Microarchitectural channels}\label{s:attack-techniques}

Microarchitectural timing channels result from competition for
capacity- or bandwidth-limited hardware features that are functionally
transparent to software \citep{Ge_YCH_18}.

Capacity-limited resources include the  data and instruction caches,
these can be used to establish
high-bandwidth
channels~\citep{Hu_92,Liu_YGHL_15,Maurice_WSGGBRM_17}. However, other
microarchitectural state, such as the translation lookasid buffer
(TLB), branch predictor (BP) or
prefetcher state machines, can be used as well.
Fundamentally, the cache channel works by the
sender (intentionally or incidentally) modulating its footprint in the
cache through its execution, and the receiver
probing this footprint by systematically touching cache lines and
measuring memory latency by observing its own execution speed. Low latency
means that a line is still in the cache from an earlier access, while
high latency means that the corresponding line has been replaced by
the sender competing for cache space. Such attacks are possible where the resource is shared concurrently
(cores or hardware threads sharing a cache) or time-multiplexed
(time-sharing a core).

Side-channel attacks are similar, except that the sender
does not actively cooperate, but accesses cache lines according to its
computational needs. Thus, the attacker must 
 synchronise its attack with the victim's execution and eliminate any
noise with more advanced techniques. Side-channel attacks have been demonstrated against
the L1-D \citep{Hu_92} and L1-I caches \citep{Aciicmez_07}, the
last-level cache (LLC) \citep{Liu_YGHL_15,Irazoqui_ES_15}, the  TLB
\citep{Hund_WH_13,Gras_RBG_18} and the BP \citep{Aciicmez_GS_07}.

Interconnects of limited bandwidth can also be used for
covert channels: the sender encodes information into its bandwidth consumption, 
and the receiver senses the available
bandwidth. So far, interconnects
can only be exploited as a covert channel while the sender and receiver execute concurrently (on
different cores). Also, if only bandwidth can be used for signalling,
side channels are probably impossible to implement, none have been
demonstrated to date.

\subsection{Countermeasures}\label{s:counter} \label{s:cache-flushing}

The countermeasures must prevent interference resulting from resource 
 competition while processing secret information. 
 For bandwidth-limited interconnects, this would
require time-multiplexing the interconnect or using some hardware
mechanism to partition available bandwidth.\footnote{Intel recently introduced 
\emph{memory bandwidth allocation} (MBA) technology, which imposes
\emph{approximate} limits on the memory bandwidth available to a
core~\citep{Intel_64_IA-32:asdm2_325383}. This is a step towards
bandwidth partitioning, but the approximate enforcement is not
sufficient for preventing covert channels.}

The OS can prevent interference on stateful resources by
flushing between accesses  or by
partitioning.\footnote{In principle, it is also possible
  to prevent timing channels by denying attackers access to real time, but in
  practice this is infeasible except in extremely constrained scenarios.}

Flushing is conceptually simple (although can
be difficult in practice, as we will discuss in \autoref{s:cannot-flush}). It is only
applicable for time-multiplexed hardware; flushing cannot
prevent cross-core attacks through a shared cache. Flushing can also be
very costly in the case of large caches (LLC), as we demonstrate in \autoref{s:flush-cost}.

Partitioning by the OS is only possible where the OS has control over
how domains access the shared infrastructure. This is the case in
physically-indexed caches (generally the L2\,\(\cdots\)\,LLC), as the OS
controls the allocation of physical memory frames to domains, and thus
the physical addresses. The standard technique is
\emph{page colouring}, which makes use of the fact that in large
set-associative caches, the set-selector bits in the address overlap with the page
number. \label{s:cache_colouring}
A particular page can therefore only ever be resident in a
specific part of the cache, referred to as the ``colour'' of the page. With a page size of
\(P\), a cache of size \(S\) and associativity \(w\) has
\(S/wP\) colours.  Therefore, the OS can partition the physically-indexed cache
with coloured frames.  By building domains with disjoint colours, 
the OS can prevent them 
competing for the same cache lines~\citep{Lynch_BF_92,
  Kessler_Hill_92, Liedtke_HH_97}.

On most hardware the OS cannot colour the small L1 caches, because 
they only have a
single colour, but also because they are  generally indexed by virtual
address, which is not under OS control.  
The same applies to the other on-core state, such as the TLB and
BP. Hence, \emph{if domains share a core, these on-core caches
must be flushed on a domain switch}.

Some architectures provide hardware mechanisms for partitioning
caches. \label{s:cat-intro} For example, many Arm processors support
pinning whole sets of the L1 I- and D-caches~\citep{ARM:ARMv7}.  Prior
work has used this feature to provide a small amount of safe, on-chip
memory for storing encryption keys~\citep{Colp_ZGSdLRSW_15}. 
Similarly, Intel recently introduced a feature called \emph{cache allocation
  technology} (CAT), which supports way-based partitioning of the
LLC, and which also can be used to provide secure
memory~\citep{Liu_GYMRHL_16}.

Although such secure memory areas can be used to protect against side
channels, we believe the time protection, like memory protection, should be a \emph{mandatory}
(black-box) OS security enforcement mechanism, rather than depending
on application cooperation. Only mandatory enforcement can support confinement.

\subsection{seL4} \label{s:seL4}

seL4 is a microkernel designed for use in security- and
safety-critical systems. It features formal, machine-checked proofs that the
implementation (at the level of the executable binary) is functionally
correct against a formal model, and
that the formal model enforces integrity and confidentiality (ignoring
timing channels)~\citep{Klein_AEMSKH_14}.

Like many other security-oriented systems \citep{Bomberger_FFHLS_92, Shapiro_SF_99}, seL4 uses capabilities
\citep{Dennis_VanHorn_66} for access control: access to any object
must be authorised by an appropriate capability. seL4 takes a somewhat
extreme view of policy-mechanism separation~\citep{Levin_CCPW_75}, by
delegating all memory management to user level. After booting up, the
kernel never allocates memory;
it has no heap and uses a strictly bounded stack. Any memory that is free
after the kernel boots is handed to the initial usermode process,
dubbed the \emph{root task}, as ``\Obj{Untyped}'' (meaning unused) memory.

Memory needed by the kernel for object metadata, including page
tables, thread control blocks (TCBs) and capability storage, must be provided to
the kernel by the usermode process which creates the need for such
data. For example, if a process wants to create a new thread, it not
only has to provide memory for that thread's stack, but it also must
hand to the kernel memory for storing the TCB. This
is done by ``re-typing'' some \Obj{Untyped} memory into the \Obj{TCB}
kernel object type. While userland now holds a capability to a kernel
object, it cannot access its data directly. Instead, the capability
is the authentication token for performing system calls on the
object (e.g.\ manipulating a thread's scheduling parameters) or
destroying the object (and thereby recovering the original \Obj{Untyped} memory).

This model of memory management aids isolation.
For example, the root task
might do nothing but partition free memory into two pools, initiate
a process in each pool, giving it complete control over its pool but
no access to anything else, and then commit suicide. This system will
then remain strictly (and provably) partitioned for the rest of its
life, with no (overt) means of communication between the
partitions. Furthermore, as kernel metadata is stored in memory
provided to the kernel by userland, it is as partitioned as userland%

\section{Attacks and Defences}

\subsection{Threat scenarios}\label{s:threat}

We aim to develop general time-protection mechanisms suitable for a wide range of use
cases. To represent these, we  pick threat scenarios from
opposite ends of the spectrum (summarised in \autoref{f:threats}). If we can satisfy both, we should be able to
address many other cases as well.

\newcommand{\Bsub}[1]{\(_{\mathbf#1}\)}

\begin{figure}[t]
  \centering
  \includegraphics[width=0.5\linewidth]{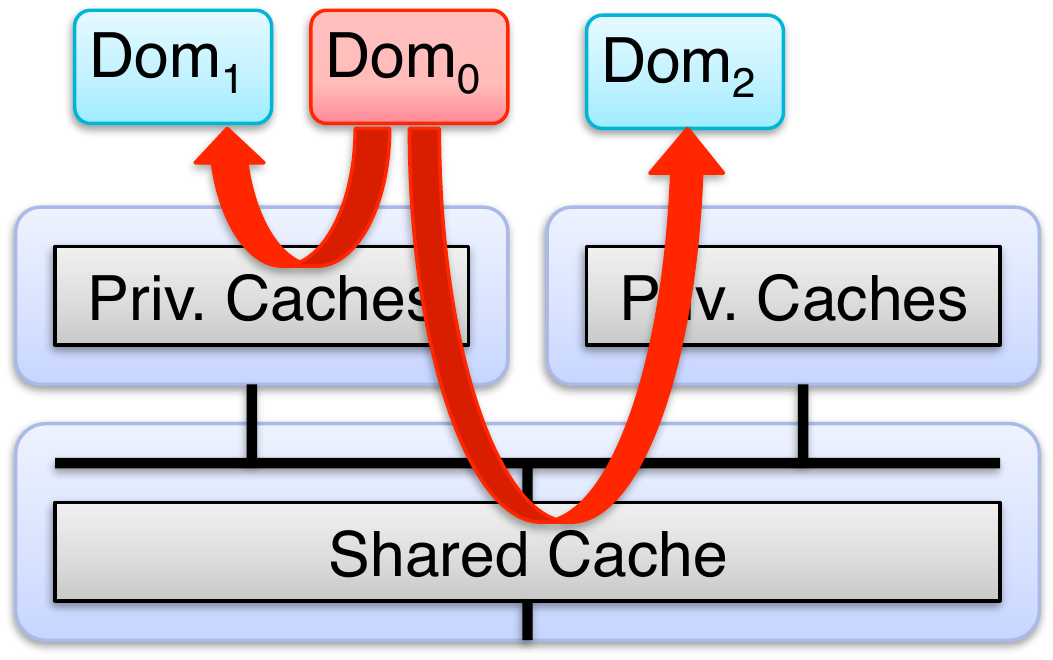}
  \caption{Threat scenarios: The arrow from Dom\Bsub{0} to
    Dom\Bsub{1}
    represents the confinement scenario of leakage through the
    intra-core covert channel, while the arrow from Dom\Bsub{0} to
    Dom\Bsub{2} indicates the cloud scenario of a cross-core side
    channel through a shared cache.}
  \label{f:threats}
\end{figure}

\subsubsection{Confinement}\label{s:confinement}

In this scenario, untrusted (malicious) code attempts to exfiltrate
sensitive data which it processes. 

This could represent untrusted code (such as a unverified library,
third-party app or web browser plugin) which has access to 
sensitive personal data, or the gadget in a Spectre attack.
An underlying assumption in these examples is that the 
code cannot be trusted not to leak information, hence the OS's time
protection must prevent leakage through a microarchitectural channel.

In this scenario we assume that the system 
either runs on a single core (at
least while the sensitive code is executing), or co-schedules domains
across the cores such that at any time only one domain executes.
We require this restriction to prevent the use of the memory
bus as a high-bandwidth channel~\citep{Hu_91,Wu_XW_12}, addressing which is
outside the scope of this work and likely impossible with current
hardware (Intel MBA notwithstanding).

\subsubsection{Cloud} \label{s:threat-cloud}

A public cloud hosts VMs
belonging to mutually-distrusting clients executing concurrently on the same processor.
As the VMs are able to communicate with the outside world, covert
channels are impossible to prevent, so the interconnect channel is of
not of much relevance. Instead we aim to prevent side channels, 
where an attacking VM is trying to infer
secrets held by a victim VM. 
Recent work demonstrated the feasibility of cross-core and cross-processor
side channel attacks through the LLC~\citep{Liu_YGHL_15,
Irazoqui_ES_15,Inci_GIES_16,Irazoqui_ES_16}. No side-channel attacks on the memory bus are known to
date~\citep{Ge_YCH_18} and they are probably
infeasible.\footnote{A recently published bus side-channel
  attack~\citep{Wang_Suh_12} was only demonstrated in a
  simulator. More importantly, it relies on the cache being small,
  making it inapplicable to modern processors.}

Multiple attacks exploiting concurrent execution within a core have
been demonstrated~\citep{Percival_05,Aciicmez_Seifert_07,Yarom_GH_16}
and hypervisor providers advise against sharing cores between VMs
\citep{Zhang_JRR_12}. 
The high level of resource sharing between hyperthreads prevents spatial
partitioning.
We therefore assume that hyperthreading is either disabled or that all
hyperthreads of a core belong to the same VM.
We do allow time-multiplexing a core between domains.

Characteristic of the cloud scenario is that it is very
performance sensitive. The business model of the cloud is
fundamentally based on maximising resource utilisation, which rules
out restrictions such as not sharing processors between VMs. This also
means that solutions that lead to significant overall performance
degradation are not acceptable.

\subsection{Requirements for time protection}\label{s:vectors}\label{s:reqts}

To address the threats above, we propose a combination of techniques
for spatially partitioning concurrently shared resources, and for flushing
time-multiplexed resources during domain switches.
As discussed in \autoref{s:cache-flushing}, flushing the virtually
indexed on-core state (L1, TLB, BP)
is unavoidable where a core is time-multiplexed between domains.
\reqbox{Flush on-core state}
{When time-sharing a core, the OS must flush on-core microarchitectural state on partition switch.\label{r:flush}}

Other core-private caches, such as the (physically addressed)  L2 in Intel processors, could
be flushed or partitioned. 
Hardware resources shared between cores, in particular the LLC, must be partitioned by the OS,
as flushing introduces too much overhead (see \autoref{s:flush-cost})
and cannot prevent timing channels in our cloud scenario.

Colouring rules out sharing of physical frames between partitions, whether explicitly or transparently via 
page deduplication, and thus may increase the aggregate memory footprint
of the system. However, this is unavoidable,
as even (read-only) sharing of code has been shown to produce
exploitable side channels~\citep{Gullasch_BK_11,Yarom_Falkner_14}. 
We are not aware of any public cloud provider
that supports cross-VM deduplication
and some hypervisor providers explicitly discourage 
the practice~\citep{VMware_KB_2080735} due to the risks it presents.

This leaves the kernel itself. Similar to shared libraries, the
kernel's code and data can also be used as a
timing channel,  we will demonstrate this in
\autoref{s:kernel-channel}.
\reqbox{Partition the kernel}
  {Each domain must have its private copy of kernel text, stack and
    (as much as possible) global data.\label{r:code}}

As discussed in \autoref{s:seL4}, partitioning most kernel data is
straightforward in seL4:
partitioning all user memory automatically partitions all dynamically
allocated (i.e.\ user-provided) kernel memory as well. Hence, colouring
user memory will colour all dynamically allocated kernel data
structures. Significantly more work would be required to implement
such partitioning in other systems, but there is no fundamental reason
why it could not be done.
This leaves a (in seL4 small) amount of global kernel data uncoloured.
\reqbox{Deterministic data sharing}
  {Access to remaining shared kernel data must be deterministic enough to prevent
    its use for timing channels.\label{r:data}}

The latency of flushing on-core caches can also be used as a channel,
as we will show in \autoref{s:switch}. The reason is that
flushing the L1-D cache forces a write-back of all dirty lines, which
means that the latency depends on the amount of dirty data, and thus
on the execution history:
\reqbox{Flush deterministically}
  {The kernel must pad cache flushing to its worst-case latency.\label{r:pad}}

Interrupts could also be used for a covert channel (although the
bandwidth would be low, as this could not signal more than a few bits
per partition switch). They are irrelevant to the cloud scenario, as 
there is no evidence that interrupts could be used as side
channels.
\reqbox{Partition interrupts}
  {When sharing a core, the kernel must disable or partition any interrupts other than the preemption timer.\label{r:irq}}

Strategies for satisfying most of these requirements are well
understood. We will now describe an approach that satisfies
\reqRef{r:code}, \reqRef{r:irq} and simplifies \reqRef{r:data} as a
side effect. Remember from \autoref{s:intro} that we are looking for mechanisms that are simple
and policy free, to make them suitable for a verifiable kernel.

\subsection{Partitioning the OS: Cloning the kernel}\label{s:clone}

\reqRef{r:code} demands per-partition copies of the kernel. It would
certainly be possible to structure a
system at boot-image configuration time, such that each partition is
given a separate kernel text segment, as in some NUMA systems~\citep{CRT:text-rep}. The partitions would still
share global kernel data, which then requires careful handling as
per \reqRef{r:data}. The latter can be simplified by reducing the
amount of shared global kernel data to a minimum, and replicate as
much of it as possible between kernel instances, resulting in
something resembling a multikernel~\citep{Baumann_BDHIPRSS_09} on a
single core, although more extreme in that kernel text is also separate.

The drawback of this approach is that it would lead to a completely static partitioning,
where the configuration of partitions, and thus the system's security
policy, is baked into the kernel boot image. As changes of policy
would require changes to the kernel itself, this reduces the degree of
assurance (or increases its cost). Especially in the case of seL4,
the initialisation code would have
to be re-verified for each configuration, or assurance lost.

We therefore favour an approach where the kernel is ignorant of the
specific security policy, only one kernel configuration (which should
eventually be completely verified) is ever used, and the security policy is
defined by the initial user process, just as with the present seL4
kernel.

We  introduce a policy-free \emph{kernel clone}  mechanism.
Its high-level description is creating a
copy of a kernel image in user-supplied memory, including a stack
and replicas of almost all global kernel data. The initial user process
can use kernel clone to set up an almost perfectly partitioned
system. Specifically, the initial process separates all free memory
into coloured pools, one per partition, clones a kernel for each partition
into memory from the partition's pool, starts a child process in each
pool, and associates the child with the corresponding kernel image,
and then commits suicide, resulting in a completely and permanently coloured system.

The existing mechanisms of seL4 are sufficient to guarantee that such
a system will remain coloured for its lifetime. Furthermore, the
process can be repeated: a partition can further sub-divide itself
with new kernel clones, as long as it has sufficient \Obj{Untyped}
memory and more than one page colour left. Partitioning can also be
reverted (assuming that the process that created it remains runnable).

\section{Implementation in seL4}\label{s:impl}

\subsection{Kernel clone overview}\label{s:clone-ov}

In seL4, all access is controlled by capabilities. To control
cloning, we introduce a new object
type,  \Obj{Kernel\_Image},  which represents a kernel. 
A holder of a \code{clone} capability to a \Obj{Kernel\_Image} object, with access to
sufficient \Obj{Untyped} memory, can clone the kernel. A
\Obj{Kernel\_Image} can be destroyed like any other object, and revoking a
\Obj{Kernel\_Image} capability destroys all kernels cloned from it.

We introduce a second new object type,
\Obj{Kernel\_Memory}, which represents physical memory that
can be mapped to a kernel image. This is analogous to the existing \Obj{Frame}
type, which represents memory that can be mapped into a user address space.

At boot time, the kernel creates a \Obj{Kernel\_Image} master
capability, which represents the present (and only) kernel and includes
the \code{clone} right. It  hands
this capability, together with the size of the image, to the initial user thread. That
thread can then partition the system by first partitioning its \Obj{Untyped}
memory by colour. For each partition it
clones a new kernel from the initial one, using some of the
partition's memory, sets up an initial address space and thread in each of
them, associates the threads with the respective kernels, and makes
them runnable. The initial
thread can prevent other threads from cloning kernels by handing them
only derived \Obj{Kernel\_Image} capabilities with the \code{clone}
right stripped.

Cloning consists of three steps. (1) The user thread retypes some
\Obj{Untyped} into an (uninitialised) \Obj{Kernel\_Image} and \Obj{Kernel\_Memory} of sufficient size, (2) it allocates an address space identifier (ASID) 
to the uninitialised \Obj{Kernel\_Image}, 
 (3) it
invokes \code{Kernel\_Clone} on the \Obj{Kernel\_Image}, passing
a  \Obj{Kernel\_Image} with \code{clone} right and
\Obj{Kernel\_Memory}   capabilities as parameters, resulting in an
initialised \Obj{Kernel\_Image}. 

Cloning copies the source kernel's code, read-only data (incl.\
interrupt vector table etc.)\ and stack. It
also creates a new idle thread and a new kernel address space; 
the seL4 kernel has an address space that contains the kernel
objects resulting from retype operations. This means that the
\Obj{Kernel\_Image} is represented as the
root of the kernel's address space, plus an ASID. Hence, any cloned \Obj{Kernel\_Image}
can independently handle any system calls, receive interrupts
(\autoref{s:interrupt-masking}) and system timer ticks, and run an
idle thread when no user thread is runnable on a core. 

The new kernel shares only the following
static data with the source kernel:
\begin{itemize}
\item the scheduler's array of head pointers to per-priority ready queues,
  as well as the bitmap used to find the highest-priority thread in
  constant time
 \item the current scheduling decision
\item the IRQ state table  and capabilities for IRQ endpoints (i.e.\ references to interrupt handlers)
\item the interrupt currently being handled (if any)
\item the first-level hardware ASID table
\item the I\/O port control table (x86)
\item the pointers for the current thread, its
  capability store (Cspace), the current kernel,  idle thread, and the thread currently owning the  floating point unit (FPU)
\item the kernel lock (for SMP)
\item the barrier used for inter processor interrupts  (SMP).
\end{itemize}

We perform an audit of the shared data to
ensure it cannot be used as a cross-core side channel.

We add to each TCB the \Obj{Kernel\_Image} capability of the
kernel that handles that thread's system calls. 
The creator of a TCB can use the
\code{TCB\_Config} system call associate the thread with a specific
\Obj{Kernel\_Image} .

\subsection{Partitioning interrupts}\label{s:interrupt-masking}

To support \reqRef{r:irq} we assign interrupt sources to a
\Obj{Kernel\_Image}. Interrupts (other than the kernel's preemption
timer) are controlled by \Obj{IRQ\_Handler} capabilities; the
\code{Kernel\_SetInt} system call allows associating an IRQ with a
kernel. 
At any time, only the preemption timer and interrupts
associated with the current \Obj{Kernel\_Image} can be unmasked, thus
preventing kernels from triggering interrupts across partition
boundaries, as long as all interrupts are partitioned. Note that policy-freedom
implies that the system will not \emph{enforce} IRQ partitioning.

\subsection{Domain-switch actions}\label{s:pad}\label{s:core-flush}

The running kernel is mostly unaware of
partitioning. As the kernel is mapped at a fixed address in the
virtual address space, the kernel (code and static data) switch
happens implicitly when switching the page-directory pointer, the only
explicit action needed for completing the kernel switch is switching
the stack (after copying the present stack to the new one). The kernel detects the need for a stack switch by comparing
the \Obj{Kernel\_Image} reference in the destination thread's TCB with
itself. 
In a properly partitioned system, the stack switch only happens on a preemption-timer interrupt. In addition, the stack switch also implies actions
for satisfying Requirements~\ref{r:flush}, \ref{r:data}, \ref{r:pad}
and \ref{r:irq}. 

We flush all on-core microarchitectural state (\reqRef{r:flush}) after 
switching stacks. The multicore
version of seL4 presently uses a big lock for performance and
verifiability~\citep{Peters_DEH_15}; we release the lock before
flushing.

To reset on-core state on Arm, we  flush the L1 caches
(\texttt{DCCISW} and \texttt{ICIALLU}), TLBs (\texttt{TLBIALL}),
and BP (\texttt{BPIALL}). On x86 we flush the TLBs
(\texttt{invpcid}) and use the recently added
\emph{indirect branch control} (IBC) feature~\citep{Intel_speculative_18}
for flushing the BP. Flushing the L1-D and -I caches presents a
challenge on x86. \label{s:cannot-flush}
While it has an
instruction for flushing the complete cache hierarchy, \texttt{wbinvd}, it has no
instruction for selectively flushing the L1 caches.
We therefore have
to implement a ``manual'' flush: The kernel sequentially traverses a
buffer the size of the L1-D cache, performing a load operation on one
word per cache line. Similarly, the kernel flushes the L1-I cache by
following a sequence of jumps through a cache-sized
buffer, which also indirectly flushes the branch target buffer (BTB).\footnote{This ``manual'' flush is obviously dependent on
  assumptions on the (undocumented) line replacement policy implemented by the
hardware, making it a brittle and potentially incomplete mechanism. Intel recently added support for 
flushing the L1-D cache~\cite{Intel_l1tf_18}. However, we cannot use
this feature, as a microcode update is yet to be available for our
machine, and there is still no L1-I flush.}

For addressing \reqRef{r:pad}, 
an authorised thread (e.g. the cloner)
may configure a switching latency. The kernel defers  returning to user mode
until the configured time is elapsed since the preemption interrupt.

Satisfying \reqRef{r:data} is much simplified by cloning, as the
kernels share almost no data (\autoref{s:clone-ov}).
We achieve determinism by carefully
prefetching all shared data before returning to
userland, by touching each cache line. This is done just prior to the
padding of the domain-switch latency. As the kernel image and stack are already switched,
and the kernel exit code path is deterministic, this prevents the
latency of the exit code from depending on the previous  domain's
execution (via lower-level caches).

To  satisfy \reqRef{r:irq}, we mask all interrupts before switching the
kernel stack, and after switching unmask the ones associated with the
new kernel.
On x86, interrupts are controlled 
by a hierarchical interrupt routing structure, all the bottom-layer
interrupts are eventually routed to the interrupt controllers on CPU
cores.  Because the kernel executes with interrupts disabled, there
exists a race condition, where an interrupt is still accepted by the
CPU just after the bottom-level IRQ source has been masked off. The
kernel resolves this by probing any possible pending interrupts 
after masking, acknowledging them at the hardware level.
Arm systems have a much simpler,
single-level interrupt control mechanism, which avoids this race.

Another race is caused by timer interrupt handling being delayed due
to another interrupt occurring just before the preemption timer. We
handle this by adding a margin for interrupt handling to the padding time.

In summary, the kernel executes the following steps when handling a preemption tick; steps in bold are only performed on a kernel switch.
\begin{enumerate} \label{s:switching-perform}
\item acquire the kernel lock
\item process the timer tick normally
\item \textbf{mask interrupts of the previous kernel}
\item \textbf{switch the kernel stack}
\item conduct the domain switch by switching the user thread (and thus the kernel image)
\item release the kernel lock
\item \textbf{unmask interrupts of this kernel}
\item \textbf{flush on-core microarchitectural state}
\item \textbf{pre-fetch shared kernel data}
\item \textbf{poll the cycle counter for the configured latency}
\item \textbf{reprogram the timer interrupt}
\item restore the user stack pointer and return.
\end{enumerate}

\subsection{Kernel destruction}

Destroying a kernel in a multicore system creates a race condition, as
the kernel that is being destroyed may be active on other cores.
For safe destruction, we first suspend all threads
belonging to the target kernel. We support this with a bitmap in each 
kernel that indicates the cores on which the kernel is presently
running, the bitmap is updated during each kernel switch.

During \Obj{Kernel\_Image} destruction, the kernel first invalidates the target
kernel capability (turning the kernel into a ``zombie''
object). It then triggers a \code{system\_stall} event, which
sends IPIs to all cores where the zombie is presently running; this is
analogous to TLB shoot-down. The cores then schedule
the idle thread belonging to the default \Obj{Kernel\_Image} (created at boot time). Similarly, the kernel sends a \code{TLB\_invalidate}  IPI
to all the cores that the target kernel had been running on. 
Lastly,
the initial core completes the destruction and cleanup of the zombie.

Destroying active \Obj{Kernel\_Memory} also invalidates the kernel,
resulting in the same sequence of actions. Destroying either object
invalidates the kernel, allowing the
remaining object to be destroyed without complications.
  
The existence of an always runnable idle thread is a
core invariant of seL4; we must maintain this invariant in the
face of kernels being dynamically created and destroyed. 
To always keep the initial kernel, we prevent the destruction of 
its \Obj{Kernel\_Memory} capability by not providing it to userland.
That way, even if userland destroys the last
\Obj{Kernel\_Image}, we guarantee that there is still a kernel and an
idle thread. Such a system will have no user-level threads, and will
do nothing more than acknowledging  timer ticks. 

One could think of more sophisticated schemes that allow reusing the
initial kernel's memory where the intention is to have a system that
is partitioned for its lifetime. For now we
accept a small amount of dead memory. On x86, where the kernel image
includes 64\,KiB of buffers used to flush the L1 caches, this
presently amounts to about 216\,KiB of waste on a single core or 300\,KiB on a 4-core machine.
Corresponding Arm sizes are 120\,KiB and 168\,KiB.

\section{Evaluation}\label{s:evaluation}

We evaluate our approach in terms of its ability to close timing
channels, as well as its effect on system performance.

\subsection{Methodology} \label{s:methodology}

For quantitative analysis of timing channels, 
we use \emph{mutual information} (MI), defined in Shannon information theory~\citep{Shannon_48}, as a measure of the size of a channel.
We model the channel as a pipe into which the sender places \emph{inputs},
drawn from some input set $I$ (the secret values), and receives 
\emph{outputs} from some set $O$ (the publicly observable time measurements).
In the case of a cache attack, the input could be the
number of cache sets the sender accesses and the output is the time it takes
the receiver to access a previously-cached buffer.
MI indicates the average number of bits of information that
a computationally unbounded receiver can learn from each input by observing the output.

We model the output time measurements as a probability density function, meaning that we are calculating the MI between discrete inputs and continuous outputs. If we treated the output time measurements as purely discrete then we would be treating all values as unordered and equivalent, e.g. a collection of unique particularly high values would not be treated differently from a collection of unique uniformly distributed values, therefore we might miss a leak. Furthermore, for a uniform input distribution, if continuous MI  is zero then it implies that other similar measures, such as discrete capacity~\citep{Shannon_48}, are also zero. As it is an average function, rather than a maximum, MI is also easier to reliably estimate, making it an effective metric to see if a leak exists or not.

We send a large number of inputs and collect the corresponding outputs. From this we use kernel density estimation~\citep{Silverman:desda} to estimate the probability density function of outputs for each input. 
Then, we use the rectangle method (see e.g.~\citep{Hughes-Hallet_GG+:calc} p. 340) to estimate the MI between a uniform distribution on inputs and the observed outputs, which we write as $\calM$. 
 
Sampling introduces noise, which will result in an apparent
non-zero MI even when no channel exists. Sampled data can never prove that a leak does not exist, so instead we ask if the data collected contains any evidence of an information leak. If  $\calM$ is very small, e.g., less that 1 millibit, we can safely say that any channel is closed or negligible. If the estimated leakage is higher than this we use the following test~\citep{Chothia_Guha_11, Chothia_KN_13} to distinguish noise in the sampling process from a significant leak. 

We simulate the
measurement noise of a zero-leakage channel by shuffling the outputs
in our dataset to randomly chosen inputs. This produces a dataset
with the same range of values, but the random assignment ensures
that there is no relation between the inputs and outputs (i.e., zero leakage). We calculate
the MI from this new dataset and repeat 100 times, giving us 100 estimations from channels that are guaranteed to have zero leakage. From this we calculate the mean and standard deviation of these results, and then calculate the exact 95\% confidence interval for an estimate to be compatible with zero leakage, which we write as $\calM_0$ (we note that the 95th highest result from the tests would only approximate the 95\% confidence interval, not give it exactly).

If the estimate of MI from the original dataset is outside the 95\% confidence interval for zero leakage (i.e., $\calM > \calM_0$) we say that the observations are inconsistent with the MI being zero, and so there is a leak (the strict inequality is important here, because for very uniform data with no leakage $\calM$ may equal $\calM_0$). If the estimated MI is within, or equal to, the 95\% confidence interval we conclude that the dataset does not contain evidence of an information leak.

\begin{table}[b]
\centering
\begin{tabular}{| l | l | l | l |l|}
\hline 
System &  Haswell (x86)  &  Sabre (ARMv7) \\
           \hline
               Microarchitecture & Haswell & Cortex A9  \\
               Processor/SoC & Core i7-4700 & i.MX 6Q \\
               Cores \(\times\) threads & \(4\times2\) &  \(4\times1\) \\
               Clock & 3.4\,GHz & 0.8\,GHz  \\
               Cache line size & 64\,B & 32\,B  \\ 
               L1-D/L1-I cache & 32\,KiB, 8-way  & 32\,KiB,  4-way\\
               L2 cache & 256\,KiB, 8-way & 1\,MiB, 16-way \\
               L3 cache & 8\,MiB, 16-way   & N/A  \\
              I-TLB & 64, 8-way  & 32, 1-way  \\
              D-TLB & 64, 4-way  & 32, 1-way  \\
              L2-TLB &  1024, 8-way & 128, 2-way  \\       
               RAM & 16\,GiB & 1\,GiB  \\                     
\hline
\end{tabular}
  \caption{\label{t:platforms} Hardware platforms.}
\end{table}

\subsection{Hardware platforms}
\label{s:hw-spec} \label{s:flush-cost}\label{s:reset-hardare-state}\label{s:scenarios}

We conduct our experiments on representatives of the x86 and Arm architectures; 
\autoref{t:platforms} gives the details.\label{s:x86plat}\label{s:armplat}
We evaluate leakage in three scenarios: \textbf{raw} refers to the
unmitigated channel while \textbf{protected} refers to
our implementation of time protection, using two coloured domains with
cloned kernels, each is allocated 50\% of available colours unless
stated otherwise.

For intra-core channels we additionally evaluate
\textbf{full flush}, which performs a maximal architecture-supported reset of
microarchitectural state. (This scenario makes no sense on inter-core
channels due to the concurrent access.)
On Arm, this adds flushing the L2 cache to the
flush operations used for time protection (as described in
\autoref{s:core-flush}), and we also disable the BP and
prefetcher for prohibiting any uncontrollable microarchitecture state. On x86 the
\textbf{full flush} scenario omits the ``manual'' L1 cache flush and
instead flushes the whole cache hierarchy (\texttt{wbinvd}), and disables
the data prefetcher by updating \texttt{MSR}~\texttt{0x1A4}~\citep{Viswanathan_14}.

As a base line we measure the worst-case direct and
indirect costs of flushing the (uncolourable) L1-I/D caches vs.\ the
complete cache hierarchy. The direct cost is the combined latency of
the flush instructions when all D-cache lines are dirty (or the cost
of the ``manual flush'' on x86). We measure the indirect cost as the one-off slowdown
experienced by an application whose working set equals the size of the
L1-D or LLC. Note that for the L1 caches, the indirect cost is
somewhat academic: It would be highly
unusual for a process to find any hot data in the L1 after another
partition has been executing for a full time slice (i.e.\ many
milliseconds).

\begin{table}[b]
  \centering
  \begin{tabular}{| l | rrr |rrr |}
    \hline
    &\multicolumn{3}{c|}{x86}&\multicolumn{3}{c|}{Arm}\\
    Cache&dir&ind&total&dir&ind&total \\
    \hline
    L1 ($\mu$s) & $25.52$ & $1.08$ & $26.59$ & $20$ & $24.53$ & $44.53$ \\
    all (ms) & $0.27$ & $0.25$ & $0.52$ & $0.38$ & $0.77$ & $1.15$ \\
    \hline
  \end{tabular}
  \caption{Worst-case cost of cache flushes.}
  \label{t:flush}
\end{table}

\autoref{t:flush} shows results. The surprisingly high L1-flush
cost on x86 is a result of our ``manual'' flush: less than
0.5\,\(\mu\)s is for the L1-D flush, the rest is for the L1-I, where
each of the chained jumps is mis-predicted. Actual flush
instructions should reduce the overall L1 flush cost to well below \(1\mu\)s.

To put these figures into context, consider that cache flushes would
only be required on a timer tick, which is typically in the order of
10--100\,ms. Flushing the L1 can be expected to add well below 1\%
overhead, while flushing the whole cache hierarchy will add
substantial overheads.

\subsection{Timing channel mitigation efficacy}
 \label{s:timing-chnannel-benchmarks}

To cover the attack scenarios listed in \autoref{s:threat}, we 
demonstrate a covert timing channel with 
a shared kernel image (\autoref{s:kernel-channel}), intra-core and 
inter-core timing channel benchmarks that exploit conflicts on 
all levels of caches (\autoref{s:cache-based-timing-channel}),  
and a timing channel based on domain switching  latency (\autoref{s:switch}).

\subsubsection{Timing channel via a shared kernel image} \label{s:kernel-channel}

\begin{figure}[t]
	\includegraphics[width=\linewidth]{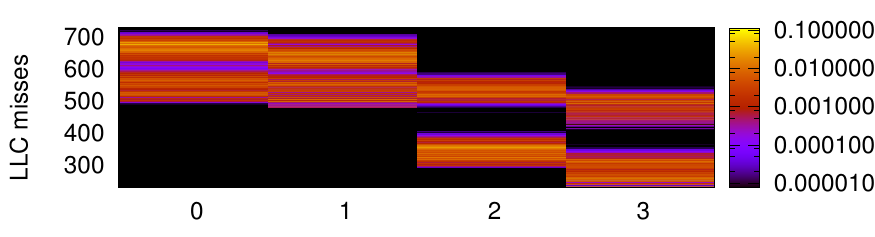}
	\includegraphics[width=\linewidth]{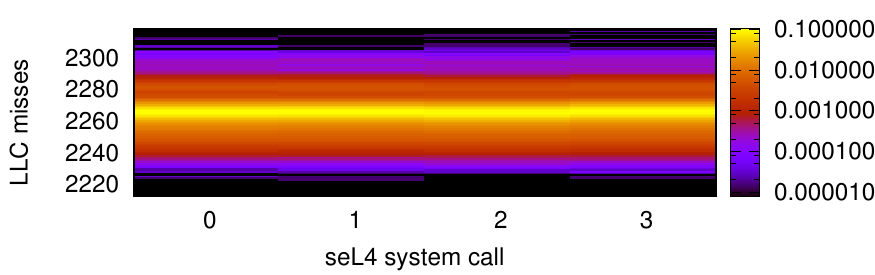}
	\caption{Kernel timing channel  on x86, with coloured
          userland (top) and full time protection (bottom). For the
          former, observed MI from a sample size
		of 255,790 is 0.79 bits, which we write as
                \stats{255,790}{0.79}. With full time protection we
                get \statsmM{255,040}{0.6}{0.1} (1\,mb = \(10^{-3}\)b).
		\label{f:cm-hw-kernel}}
\end{figure}

We demonstrate that partitioning only user space is insufficient
for mitigating covert channels, even though on seL4 this automatically
partitions dynamic kernel data (\autoref{s:seL4}). Among others, this
setup already defeats the attack of \citet{Schaik_GBR_18}, as page tables are
automatically coloured.

We implement an LLC  covert channel between coloured user-space
processes. The sender sends information  by triggering system calls,
while the receiver, sharing the same core with a time slice of 1\,ms, monitors the cache misses on the cache set that kernel uses for serving the system calls.

The receiver firstly builds a probe buffer with the \pp technique~\citep{Osvik_ST_06,Percival_05,Liu_YGHL_15}: 
it compares the cache misses on the probed cache sets before and after executing the system call, 
then marks a cache set as an attack set if the number of cache misses increase after the 
system call is returned. 

The sender encodes a random sequence of symbols from the set $I={0,1,2,3}$
by using  three system calls:  \code{Signal} for 0,
\code{TCB\_SetPriority} for  1, \code{Poll} for  2, and idling
for 3. \autoref{f:cm-hw-kernel} (top) shows the resulting \emph{channel
  matrix}, which represents the conditional probability of observing
an output symbol given a particular input symbol, shown as a heat
map. A channel is indicated by output symbols (cache misses) being correlated with
input symbols (system calls), i.e.\ variations of probability (colour)
along horizontal lines. \code{Signal} and
\code{TCB\_SetPriority}  lead to  500--700 misses, while \code{Poll}
and idle result in 200--600 misses, a clear channel. Calculating the MI  gives an
estimated information leak of  $\calM$=0.79 bit per iteration (2\,ms),
transmitting 395\,b/s. While the channel could be made more efficient
with more complicated encoding schemes, this is not the main focus of our work.

With cloned kernels the channels disappear
(bottom of \autoref{f:cm-hw-kernel}). The remaining channel is measured as
$\calM=0.6$ millibits (mb), therefore closed or negligible.
We implement a similar channel on the Arm, observing a non-trivial 
MI $\calM=20$ mb, which reduces to $\calM=0.0$\,mb with time protection.

\begin{table}[b]\centering
	\begin{tabular}{|l|l|r|r@{ }l|r@{ }l|}
		\hline
		\bf Platform & \bf Cache & \bf Raw & \multicolumn{2}{c|}{\bf Full flush}& \multicolumn{2}{c|}{\bf Protected}\\
		\hline
		x86
		&L1-D & 4,000 & \hspace*{0.6em}0.5 & (0.5) & \hspace*{0.4em}0.6  & (0.6) \\ 
		&L1-I & 300 & 0.7 & (0.8) & 0.8 & (0.5) \\ 
		&TLB & 2,300 & 0.5 & (0.5) & 16.8 & (23.9) \\ 
		&BTB & 1,500 & 0.8 & (0.8) & 0.4 &  (0.4)\\ 
		&BHB & 1,000 & 0.5  & (0.0) & 0.0  & (0.0) \\ 
		&L2 & 2,700 & 2.3 & (2.6) & 50.5 &  (3.7) \\
		\hline
		Arm
		&L1-D & 2,000 & 1 & (1) & 30.2 & (39.7)\\ 
		&L1-I & 2,500 & 1.3 & (1.3) & 4.9 & (5.2) \\ 
		&TLB & 600 & 0.5 & (0.5) & 1.9 & (2.2)\\ 
		&BTB & 7.5 & 4.1 & (4.4) & 62.2 & (73.5) \\ 
		&BHB & 1,000 & 0 & (0.5) & 0.2 & (54.4) \\ 
		&L2 & 1,900 & 21 & (22) & 1.4 &  (1.4)\\ 
		\hline
	\end{tabular}	
	\caption{MI  (mb) of unmitigated (raw)
		intra-core timing channels, mitigated with full
		cache flush (full flush)  and  time protection (protected).
		Value in parentheses is $\calM_0$. \label{t:cache-channel-result}}
\end{table}

\subsubsection{Intra-core timing channels} \label{s:cache-based-timing-channel}

We investigate the a full set of channels exploitable by processes
time-sharing a core, targeting the L1-I, L1-D and L2 caches, the TLB,
the  BTB,  and the branch history buffer (BHB). 
 We use a \pp attack, where the receiver measures the timing on  probing a defined number of 
 cache sets or entries.

We use the Mastik~\citep{Yarom_16:Mastik} implementation of the \textbf{L1-D}
cache channel, the output symbol is the time to perform the attack on every
cache set. The \textbf{L2} channel is the same with a probing set
large enough to cover that cache. We build the \textbf{L1-I} channel by 
having the sender probe with jump instructions that 
map to corresponding cache sets~\citep{Aciicmez_07,
  Aciicmez_BG_10}. For the \textbf{TLB} channel, the sender probes the TLB entries by 
reading a integer from a number of consecutive pages.  We use a chained branch instructions as the probing 
buffer for the \textbf{BTB} channel,  the sender probing 3584--3712 branch
instructions on Haswell,  0--512 on Sabre. Our \textbf{BHB} channel is the same as the residual state-based covert channel~\citep{Evtyushkin_PA_16:TACO}, where the sender sends 
information by either taking or skiping a conditional jump instruction. 
The receiver measures the  latency on a similar conditional jump instruction, sensing 
any speculative execution caused by the sender's history. 

\autoref{t:cache-channel-result} summarises results for the three scenarios defined in
\autoref{s:scenarios}. The \emph{raw} scenario shows a large channel
in each case. On the Sabre we find that all channels are effectively
mitigated by the \emph{full flush} as well as the \emph{protected
  scenario}.

On Haswell, the picture is the same except for the L1-I and L2
channels. The L1-I channel does not seem quite closed with time
protection, although the
residual capacity is negligible, and likely results from our imperfect
``manual'' flush. While the full flush closes the L2 channel, our
implementation of time protection (which colours the L2) fails to do this, leaving a sizeable
channel of 50\,mb. We believe 
that the remaining channel is due to the aggressive data prefetcher, 
as the channel is decreased to $\calM=6.4$ mb ($\calM_0=4.1$ mb) 
with the data prefetcher disabled. This is strong 
evidence for the need of a better software-hardware contract for controlling  
any hidden microarchitecture state~\citep{Ge_YH_18}.

\begin{figure}[t]
	\includegraphics[width=\linewidth]{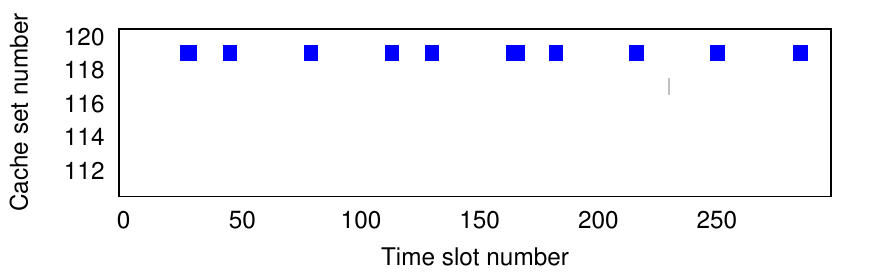}
	\caption{Unmitigated concurrent LLC side-channel attack on x86.  The pattern in blue shows
		victim's cache footprint detected by the spy.
		\label{f:side-haswell-llc}}
\end{figure}

\subsubsection{Side channel on the LLC} \label{p:side-channel-llc}
To test the side channel mitigation for LLC-based attacks,
we reproduce the attack of \citet{Liu_YGHL_15} on GnuPG
version 1.4.13.
The attack targets the square-and-multiply implementation of
modular exponentiation used as part of the ElGamal decryption.

\begin{table}[b]
	\begin{tabular}{|l|c|r|c|}
		\hline
		\bf Platform & \bf Timing & \bf No pad & \bf Protected ($\calM_0$) \\
          \hline
          \bf x86
		& On-line &  8.4  & 0.5 (0.5)\\ 
          pad = 58.8 $\mu$s
		& Off-line & 8.3 & 0.6 (0.6) \\ 
          \hline
          \bf Arm
	    & On-line & 1,400  & 16.3 (24.6)\\ 
          pad = 62.5 $\mu$s
		& Off-line & 1,400 & 210 (237.2)  \\ 
          \hline
	\end{tabular}
	\caption{Channel resulting from cache-flush latency (mb)
          without and with time protection.
	 \label{t:cache-latency-result}}
\end{table}

We use two  processes, executing concurrently on separate cores 
on Haswell. 
The victim repeatedly decrypts a file, whereas the spy uses the Mastik implementation of the LLC \pp
attack to capture the victim's cache activity, searching for patterns that correspond 
to the use of the square function. The cache activity learned by the spy is shown on \autoref{f:side-haswell-llc}. On cache set
number 119, we see a sequence of blue dots separated
by intervals of varying lengths.  
Each of these dots is an invocation of the square function
and the secret key is encoded in the length of the intervals
between the dots, with  long intervals encoding ones and
short intervals zeros.
We find that time protection closes the channel (in this case by
colouring the LLC),
the spy can no longer detect any cache activity of the victim.

\subsubsection{Cache-flush channel\label{s:switch}}

To demonstrate the cache-flush channel we create a receiver that
measures two properties: \emph{online time} is the receiver's
uninterrupted execution time (i.e.\ the observed length of its time
slice) while \emph{offline time} as the time between receiver's
executions. The receiver repeatedly checks the  cycle counter, 
waiting on a large  jump that  indicates a preemption.
The length of the jump is the offline time, whereas the 
 time between consecutive intervals is the online time. 

\begin{figure}[t]
	\includegraphics[width=1.05\linewidth]{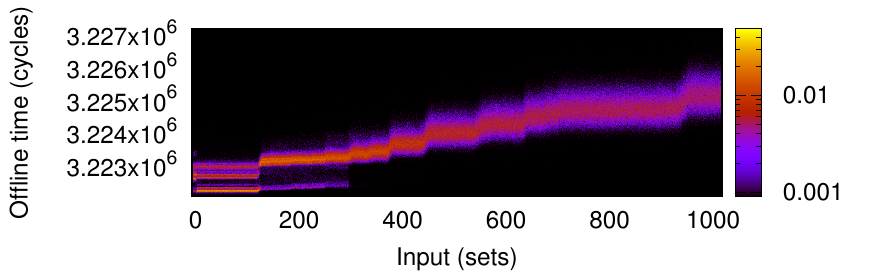}
	\caption{The unmitigated offline time observed by the receiver vs.\ the
          sender's cache footprint, resulting in a variation of domain switching 
          latency on Sabre.
		\stats{1828}{1.4}
		\label{f:cm-sabre-kd-none}}
\end{figure}

The sender  varies the number of L2 cache
sets it accesses in each time slice, 
manipulating  the cost of the kernel's L1 cache flushes, and thus the
receiver's online or offline time.

\autoref{f:cm-sabre-kd-none} shows that the sender effectively
modulates the offline time. 
\autoref{t:cache-latency-result} shows that the channel exists on both
architectures, but is effectively closed with time padding.

\subsubsection{Interrupt channel}\label{s:controlled-interrupt-delivery}

\begin{table}[b]
	\begin{tabular}{|l|c|r@{~}l|r@{~}l|}
		\hline
		\bf Platform & \bf IRQs & \multicolumn{2}{c|}{\bf Shared} & \multicolumn{2}{c|}{\bf Partitioned}\\
		\hline
		\bf x86
	&	no &  \hspace*{0.5em}10&(0)  & \hspace*{1.5em}10&(0)\\ 
	&	yes & 5&(5)& 10&(0) \\ 
		\hline
		\bf Arm
	&	no &  10&(0)  & 10&(0)\\ 
&	yes & 5 &(5) & 10&(0) \\ 
		\hline
	\end{tabular}
	\caption{Interrupt channel vs.\ IRQ partitioning: receiver
          on-line time in ms (standard deviation).
		\label{t:interrupt-partition-result}}
\end{table}
We evaluate interrupt partitioning with a sender which sends
``no'' by doing nothing or ``yes'' by  programming a timer to fire
every millisecond, the receiver is as in \autoref{s:switch}. Without interrupt
partitioning, if the interrupt fires while the receiver is executing,
in average 0.5\,ms into its time slice, the kernel will be invoked,
resulting in the receiver recording a short on-line time. As the
kernel has masked the IRQ, it will not fire again until the sender
acknowledges it, and the receiver will not be interrupted a second
time, and thus record a long on-line time, in average 9.5\,ms for a
10m\,ms time slice. This bi-modal distribution is an effective channel
and is reflected in the large, 5\,ms standard deviation in
\autoref{t:interrupt-partition-result}.

With interrupt partitioning, the receiver is not preempted during its
time slice, resulting in a deterministic on-line time, and thus a
closed channel.

\subsection{Performance} \label{s:system-benchmarks}

\subsubsection{IPC microbenchmarks} \label{s:ipc}

We evaluate the impact of time protection by measuring the cost of the most important
(and highly optimised) microkernel operation, cross-address-space message-passing IPC.
\autoref{t:ipc} summarises the results, where \emph{Colour ready} refers to
a kernel supporting time protection without using it, intra-colour
measures IPC that does not cross domains (kernels), while inter-colour
does. The last is an artificial case that does not use a fixed time
slice or time padding (which would defer IPC delivery to the partition switch)
examining baseline cost of our mechanisms. 
Standard deviations from 30
runs are less than 1\%.

\begin{table}[t]
\begin{center}
\centering
\begin{tabular}{|l|cc|cc|}
\hline
	& \multicolumn{2}{c|}{\bf x86}	& \multicolumn{2}{c|}{\bf Arm}\\
  \bf Version & \bf Cycles & \bf O/H & \bf Cycles & \bf O/H\\
  \hline
  original &  381 & - &  344 & - \\
  colour-ready & 386 & 1\% & 391 & 14\% \\ 
  intra-colour & 380 & 0\% & 395 & 15\% \\
  inter-colour & 378 & -1\% & 389 & 13\% \\ 
  \hline
\end{tabular}
\caption{\label{t:ipc} IPC performance microbenchmarks.}
\end{center}
\end{table}

We find that the time-protection mechanisms add negligible overhead on
x86. On  Arm, in contrast,  there is a significant baseline cost to
supporting the kernel clone mechanism, resulting from the fact
that with multiple kernels, we can no longer use global mappings with large entries to
map the kernel's virtual address space. The Sabre's A9 core has a
2-way L2-TLB, resulting in increased conflict misses on 
the cross-address-space IPC test. There is no further
overhead from using cloning.

\begin{table}[b]
	\begin{center}
		\centering
	  \begin{tabular}{|l|l|r|r|r|r|r|}
				\hline
		\bf Platf. & \bf Mode & \bf Idle  & \bf L1-D  & \bf L1-I  & \bf L2 & \bf LLC  \\
	\hline
            \bf x86
		&Raw & 0.18  &  0.19&  0.22 &  0.23 & 0.5 \\
		&Full flush & 271  & 271 & 271  & 271 & 271\\ 
		&Protected  &  30 & 30 & 30 & 30 & 30\\ 
	\hline
            \bf Arm
		&Raw & 0.7  & 0.8 &  1.2 & N/A& 1.6 \\
		&Full flush &  414 & 414 &  414 & N/A &414 \\ 
		&Protected  & 27 & 27 & 27 & N/A &31\\ 
		\hline
	\end{tabular}
		\caption{\label{t:context-switch-cost} Cost  ($\mu$s) of
                  switching away from a domain running various
                  receivers from \autoref{s:cache-based-timing-channel}.}
	\end{center}
\end{table}

\subsubsection{Domain switching cost} \label{s:domain-switch-cost}

In \autoref{t:flush} we measured the worst-case cache-flush costs. We
expect those to dominate the cost added to domain switches by time
protection. To verify we evaluate the domain-switch latency
(without padding) for a number of our attack workloads. Specifically
we measure the time taken to switch from the receiver of a \pp attack
to an idle domain. We report the mean for 320 runs, all
standard deviations are less than 1\% (ARM) or 3\% (x86).
An exception is the LLC test, where original seL4 times have a bimodal
distribution and  we report  median values (standard deviation: 25\% for Arm, 18\% for x86 ).

\autoref{t:context-switch-cost} shows the results for our three
defence scenarios. We observe first that the workload dependence of
the latency evident in the raw system has mostly vanished from the
defended systems, even without padding. We second notice that, as
expected, the \emph{full flush} latencies match the flush costs of
\autoref{t:flush}. With time protection, the switch latency is
slightly higher than the direct L1-flush cost of \autoref{t:flush},
confirming our hypothesis that this is the dominant cost, and also
supporting the comment in \autoref{s:flush-cost} that indirect flush
cost are of little relevance for L1 caches.

Most importantly, the results show that our implementation of time
protection imposes significantly less overhead than the full flush,
despite being as effective in removing timing channels (except for the
issues resulting from the lack of targeted cache flushes discussed in
\autoref{s:cache-based-timing-channel}).

\subsubsection{The cost of cache colouring}

\begin{figure}[t]
        \includegraphics[width=\linewidth]{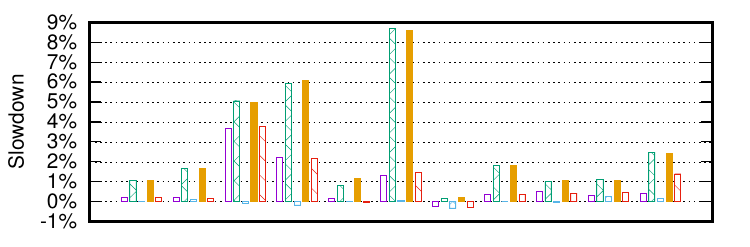}\\
	\includegraphics[width=\linewidth]{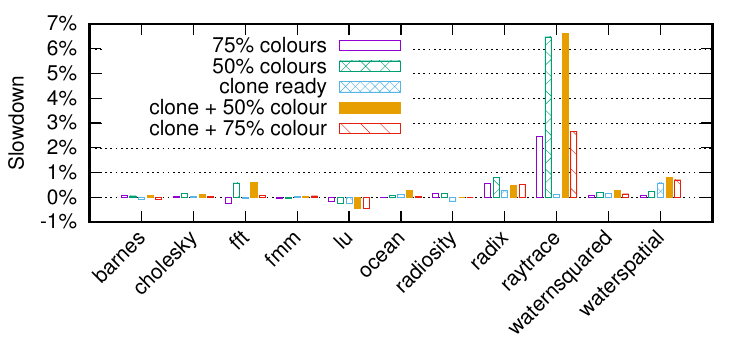}
	\caption{Splash-2 performance with cache colouring for x86
          (top) and Arm.
		\label{f:splash-colouring}}
\end{figure}

\newcommand{\splash}[1]{\textsl{#1}}

To evaluate the cost of cache colouring, we port the Splash-2 benchmark~\citep{Woo_OTSG_95} to the seL4 system (except volrend due to its Linux dependencies), with running parameters 
that consume 220\,MiB of heap and 1\,MiB of stack. \autoref{f:splash-colouring}
shows the overhead of cache colouring with and without the kernel clone mechanism. 
We report the mean of 10 repeated single-threaded runs (standard
deviations are below 3\%). The benchmarking thread 
is the only user thread in the system. 

On  Sabre, cache colouring introduces less 
than 1\% slowdown for most of the benchmarks. The only exception is
\splash{raytrace}, which shows  a  6.5\%  slowdown when executing with  
50\% of the cache, as this benchmark has a large cache working
set. However, given a 75\% cache share, the slowdown drops to
2.5\%. On top of this, running on a cloned kernel adds almost no
performance penalty, except 
on \splash{waterspatial}, where it is still below 0.5\%.  

On Haswell, we observe slightly larger performance overheads,
as we partition based on colours of the relatively small (256\,KiB) L2 cache
(which implicitly colours the LLC). The alternative would be to only colour the LLC
and flush the L2, but with no targeted L2 flush supported by the
architecture this seems not worthwhile. Still, the majority of the 
Splash-2 tests only slow down by less than 2\%. Increasing cache share
to 75\% limits the overhead to below 3.5\%. As for Arm, the kernel clone mechanism 
has close to zero overhead.

\subsubsection{The impact of domain switches}

\begin{table}[b]
	\begin{center}
		\centering
		\begin{tabular}{|l|rr|rr|}
			\hline
			& \multicolumn{2}{c|}{\bf x86}	& \multicolumn{2}{c|}{\bf Arm}\\
			\bf Version & \bf ocean & \bf radiosity & \bf lu & \bf raytrace\\
			\hline
			50\% colour                 & 4.8\%  & -0.5\% & 0.03\%   & -2.4\% \\ 
		  50\%  \texttt{+} padding & 5.5\%  & 0.1\% & 0.3\%  & -2.0\% \\ 
		  	75\% colour                  & -0.3\%  & -0.5\% & -0.02\%  & -6\% \\ 
		  75\% \texttt{+} padding & 0.4\%   & 0.1\% & 0.2\%  & -5.8\% \\ 
			\hline
		\end{tabular}
		\caption{\label{t:splash-domain-switch} Splash-2
                  performance overhead from time protection with time
                  padding disabled or enabled.}
	\end{center}
\end{table}

For evaluating the full impact of time protection, we select from
Splash-2 the benchmarks suffering the highest and lowest
cache-partitioning overheads according to \autoref{f:splash-colouring}. We
simulate a timing-channel defence scenario, with the Splash benchmark
sharing a core with an attacking thread. The latter
is continuously probing the L1-I and the LLC caches.  We use full time
protection with a 10\,ms time slice. We give the Splash program 50\%
or 75\% of the cache and use the padding times of
\autoref{t:cache-latency-result}. Note these are well above the
worst-case L1 flush costs of \autoref{t:flush} and could be
optimised.
We report in  \autoref{t:splash-domain-switch} averages of 10 runs
(standard deviations below 0.1\%).

On Haswell (x86), cache partitioning actually improves performance of
the \splash{radiocity} benchmark: it provides performance isolation, removing the
frequent conflict misses resulting from the attack thread in the
unprotected system. This performance gain offsets the increased
context-switch latency from time padding; we see the same effect on
\splash{ocean} when it gets a 75\% share of the cache. The overhead
resulting from padding is about 0.7\%.

The performance isolation effect is even more pronounced on the Arm:
\splash{raytrace} consistently performs better with a partitioned
cache, and the performance of \splash{lu} is practically unaffected by
partitioning. Padding only introduces 
0.2\%--0.4\% performance overhead.

\section{Related  Work} \label{s:related-work}

Deterministic systems eliminate timing channels by providing only
virtual time;
Determinator \cite{Aviram_HFG_10} is an
example aimed at clouds.   \citet{Ford_12} 
extends this model with scheduled IO.  Stopwatch~\citep{Li_GR_13}  visualizes time by running  three replicas of a
system, then only announces externally-visible timing events at the median of the
times determined by the replicas. The system is effective but at a
heavy performance penalty.

Page colouring for partitioning caches goes back to 
\citet{Kessler_Hill_92, Bershad_LRC_94}, who  proposed it for
performance isolation. \citet{Liedtke_HH_97} proposed the same for
improved real-time predictability, while \citet{Shi_SCZ_11} proposed
dynamic page colouring  for mitigating attacks against
cryptographic algorithms in the
hypervisor. \textsc{StealthMem}~\citep{Kim_PM_12} uses colouring to
provide some safe storage with controlled cache
residency. CATalyst~\citep{Liu_GYMRHL_16} uses Intel's CAT
technology for LLC partitioning for a similar purpose.
 
\citet{Percival_05} proposed hardware-supported partitioning of the L1
cache, while  \citet{Wang_Lee_07} suggested hardware
mechanisms for locking cache lines, called a partition-locked cache (PLcache). 
\citet{Ge_YH_18} investigate shortcomings in architectural support for
preventing timing channels and propose an extended hardware-software
contract.

Multikernels~\citep{Baumann_BDHIPRSS_09} consist of multiple,
shared-nothing kernel images on the same hardware platform, although
on separate cores, for improved many-core scalability. Barrellfish/DS~\citep{Zellweger_GKR_14} 
separates OS kernel images from physical CPU cores, to support
hot-plugging and energy management.

\section{Conclusions} \label{s:future-work}

We proposed, implemented and evaluated \emph{time protection}, a mandatory, black-box kernel mechanism for
preventing microarchitectural timing channels. Time protection employs
a combination of cache partitioning through colouring and flushing of
non-partitionable hardware state. It leverages a policy-free
\emph{kernel clone} mechanism to
almost perfectly partition the kernel itself, resulting in a
per-partition kernel image on each core, with interrupts being
partitioned as well. Our evaluation shows that the mechanisms are  effective 
for closing all studied timing channels, while imposing small to
negligible performance overhead. While present x86 hardware has some
shortcomings that prevent perfect time protection, it would be easy
for manufacturers to address this by supporting more targeted flush
operations for microarchitectural state.

\label{p:lastpage}
\balance 

\bibliographystyle{ACM-Reference-Format}
 \bibliography{paper}
\end{document}